\documentclass[12pt]{iopart}
\usepackage[dvips]{graphicx}
\newcommand{\EPL}{Europhys.~Lett.~}
\newcommand{\MP}{Mol.~Phys.~}
\newcommand{\JCIS}{J.~Coll.~Int.~Sci.~}
\newcommand{\EPJ}{Eur.~Phys.~J.~}

\begin{document}

\title[Charge Renormalization and Thermodynamics of Colloidal Suspensions]
{Charge Renormalization, Effective Interactions, and Thermodynamics
of Deionized Colloidal Suspensions}

\author{A. R. Denton\footnote[1]{Electronic address: {\tt alan.denton@ndsu.edu}}}
\address{Dept.~of Physics, North Dakota State University, Fargo, ND, U.S.A.
58108-6050}

\date{\today}

\begin{abstract}
Thermodynamic properties of charge-stabilised colloidal suspensions depend
sensitively on the effective charge of the macroions, which can be
substantially lower than the bare charge in the case of strong
counterion-macroion association.  A theory of charge renormalization is
proposed, combining an effective one-component model of charged colloids with
a thermal criterion for distinguishing between free and associated counterions.
The theory predicts, with minimal computational effort, osmotic pressures
of deionized suspensions of highly charged colloids in close agreement with
large-scale simulations of the primitive model.
\end{abstract}

\pacs{82.70.Dd, 83.70.Hq, 05.20.Jj, 05.70.-a}

\maketitle

\section{Introduction}\label{intro}

Colloidal suspensions of charged macroions --- nanometers to micrometers in
size and dispersed in a fluid by Brownian motion --- are ubiquitous in nature
and industry~\cite{Evans,Schmitz}.  The remarkable thermal, optical, and dynamical
properties of colloidal materials hinge on a delicate balance between
competing interparticle interactions~\cite{Israelachvili}.  Common examples
include aqueous paints, detergents, and clays: dispersions of latex particles,
ionic surfactant micelles, and mineral platelets, respectively.
Self-assembled crystals of synthetic, monodisperse, silica or polystyrene
microspheres provide useful scaled-up models of atomic crystals and promise
novel technologies, such as photonic band-gap materials~\cite{photonic1,photonic2}.
As predicted by the classic theory of Derjaguin, Landau, Verwey, and Overbeek
(DLVO)~\cite{DL,VO}, repulsive electrostatic forces between charged colloids
can stabilise a suspension against aggregation induced by van der Waals
attractive forces~\cite{interactions}.

Dispersed in a polar solvent, colloidal particles can acquire charge through
dissociation of ionizable chemical groups at the surface.  Electrostatic interactions
are sensitive to the surface charges of the macroions and to the distribution of
surrounding counterions.  A macroion's bare (structural) charge depends on its
surface chemistry (e.g., number and type of ionizable sites) and, in general,
on the pH and salinity of the solution~\cite{belloni91,vongrunberg99}.  Dressed by
an entourage of strongly attracted counterions, a highly charged macroion can act
as though carrying a significantly reduced (renormalized) {\it effective} charge.

The basic concepts of charge renormalization and effective charge were first
introduced and widely explored some four decades ago in the context of polyelectrolyte
solutions~\cite{manning69,Oosawa}.  Similar ideas were subsequently applied
to colloidal suspensions by Alexander {\it et al.}~\cite{alexander84},
who demonstrated that strong association of counterions can significantly
renormalize spherical macroion charges.  Numerous experimental studies of
deionized aqueous suspensions of highly charged spherical latex
particles~\cite{gisler94,palberg95,palberg02,grier00}, integral-equation
calculations~\cite{belloni86}, and simulation studies of the primitive 
model~\cite{stevens95,linse-lobaskin99,linse00,lobaskin-linse01,lobaskin-qamhieh03}
have since confirmed the effective charge as a physically important parameter
in the one-component model of colloidal suspensions.

The one-component model provides a practical approach to overcoming the severe
challenges of extreme size and charge asymmetries in explicit molecular models of
charge-stabilised colloidal suspensions, polyelectrolyte solutions, and many other
soft materials~\cite{zvelindovsky07}.  The model is derived from the multi-component
ion mixture by averaging over the degrees of freedom of the microions
(counterions and salt ions).  The surviving ``pseudo-macroions" are governed by
effective electrostatic interactions, screened by the implicitly modeled microions.

This paper seeks to unite the concepts of dressed macroions and effective
interactions in a coherent statistical mechanical framework that describes
the association of counterions with macroions, the renormalization of the
effective macroion charge, effective electrostatic interactions between
macroions, and thermodynamic properties of deionized suspensions of highly
charged colloids.  Conceptually similar syntheses have been proposed recently,
based on Debye-H\"uckel theory~\cite{levin98,tamashiro98,levin01,levin03,trizac04}
and on nonlinear Poisson-Boltzmann theory~\cite{trizac02,zoetekouw_prl06}.
The present theory is inspired by the elegant liquid-state approaches of Levin,
Trizac, and coworkers~\cite{levin98,tamashiro98,levin01,levin03,trizac04},
but differs in several significant practical respects.

The remainder of the paper is organized as follows.  Sections~\ref{model} and
\ref{theory} trace a path from the microscopic primitive model of charged colloids
to an effective one-component model of dressed, charge-renormalized macroions.
A simple criterion is adopted to differentiate between free and electrostatically
bound counterions; physical approximations are developed for the free energies
of the two counterion phases; and a variational method is prescribed for
determining the renormalized effective charge and screening constant.
Section~\ref{results} demonstrates the practical implementation of the theory and
compares predictions for the pressure of deionized suspensions with corresponding
data from both simulations of the primitive model and experiment.  Excellent
agreement is obtained, over broad ranges of system parameters, with trivial
computational effort.  Finally, Sec.~\ref{conclusions} closes with a summary and
perspectives.

\section{Model}\label{model}

Within the primitive model of charged colloids, the macroions are modeled as
negatively charged hard spheres of monodisperse radius $a$ and bare valence $Z_0$
(charge $-Z_0e$), the microions as monovalent point charges, and the solvent
as a dielectric continuum of uniform relative permittivity $\epsilon$.
Polarization effects and image charges are ignored, assuming index-matching of
macroions and solvent.
The suspension may be either entirely confined to a closed volume at fixed salt
concentration or in partial chemical (Donnan) equilibrium (e.g., via a
semi-permeable membrane) with a microion reservoir, which fixes the microion
chemical potentials.  The reservoir is presumed to be a 1:1 electrolyte solution
with number density $n_0$ of monovalent salt ion pairs.

In the presence of a sufficiently strong attractive potential, some fraction of
counterions may remain closely associated with the macroions.  By analogy with
Oosawa's two-phase theory of polyelectrolyte solutions~\cite{Oosawa}, a distinction
then can be drawn between free and bound microion regions (``phases").  In contrast
to rodlike polyelectrolytes, however, spherical colloidal macroions do not generate
a Coulomb potential of sufficient range to overcome counterion entropy and condense
the counterions.  As a result, associated counterions remain only thermally
(not physically or chemically) bound to the macroions.

\begin{figure}
\begin{center}
\includegraphics[width=0.5\columnwidth]{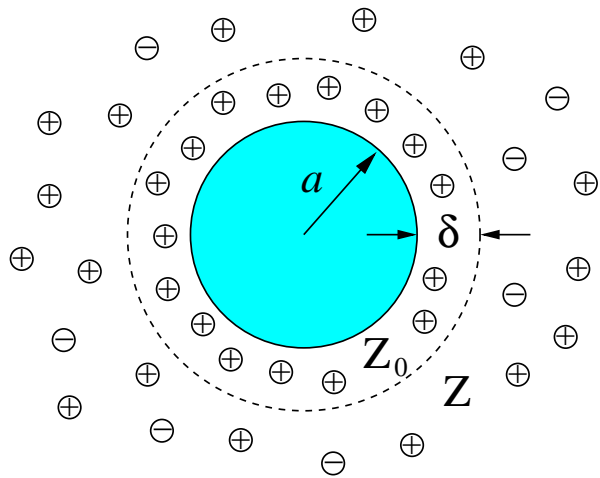}
\end{center}
\vspace*{-0.5cm}
\caption{\label{fig-model}
Model of charged colloidal suspension: spherical macroions of radius $a$
and point microions dispersed in a dielectric continuum.  Strongly associated
counterions in a spherical shell of thickness $\delta$ renormalize the
bare macroion valence $Z_0$ to an effective (lower) valence $Z$.
}
\end{figure}

As Fig.~\ref{fig-model} depicts, counterions localized within a
spherical shell of thickness $\delta$ (yet to be determined) are regarded as
renormalizing the bare macroion valence.  Coions are assumed to be completely
expelled from the shell.  The resulting ``dressed" macroion is a composite object
consisting of a bare macroion and its shell of bound counterions with an
effective valence $Z\le Z_0$.
Although the bare and effective valences are statistically fluctuating quantities,
they are represented for present purposes by their average values.

The boundary between free and bound counterions is located at a distance from
the macroion surface at which the electrostatic energy of counterion-macroion
attraction is comparable to the average thermal energy per counterion.
Denoting by $\phi(r)$ the electrostatic potential at distance $r$ from a
macroion centre, the association shell is defined via
\begin{equation}
e|\phi(a+\delta)|=Ck_BT~,
\label{delta1}
\end{equation}
where $C$ is an adjustable, dimensionless parameter, evidently of order unity.
Counterions within the association shells ($a<r<a+\delta$) are assumed to be
trapped in the potential wells of the macroions, while more distant counterions have
sufficient kinetic (thermal) energy to escape.  This simple criterion for $\delta$
justifies a Debye-H\"uckel-like linear-screening approximation for the free
counterions, which is exploited in the theory developed below.

Previous studies have applied a thermal criterion similar to Eq.~(\ref{delta1})
to the electrostatic potential~\cite{schmitz99,schmitz00,schmitz02} or to the
effective pair potential~\cite{safran00}.  Alternative approaches to defining
the association shell thickness are based on the structure of the counterions
around a macroion or on the macroion configurations.
Alexander {\it et al.}~\cite{alexander84},
for example, determined $Z$ in a spherical cell model by matching the solutions
of the nonlinear and linearized PB equations for the counterion density at the
edge of the cell.  The inflection point in the running effective charge of the
macroion has been identified as another sensible boundary between free and bound
counterions~\cite{belloni98}.  Yet another fruitful approach is to fit the effective
one-component model (with a screened-Coulomb pair potential) to either the static
structure factor measured in light-scattering experiments~\cite{gisler94} or
the pair distribution function computed in simulations of the primitive
model~\cite{stevens95,linse-lobaskin99,linse00,lobaskin-linse01,lobaskin-qamhieh03}.
A useful comparison of various criteria for defining effective charges is
provided in ref.~\cite{schmitz02}.

\section{Theory}\label{theory}

The theory proposed here for charge renormalization and thermodynamics of
colloidal suspensions requires modeling the electrostatic potential and the
total free energy of the system.  For this purpose, the most popular framework
is the Poisson-Boltzmann (PB) theory~\cite{deserno-holm01}, a
mean-field approach that is especially well-suited to suspensions with
monovalent microions, whose correlations usually can be justifiably neglected.
Combining the exact Poisson equation for the potential with a Boltzmann
approximation for the microion density profiles (as functions of position
${\bf r}$), $n_{\pm}({\bf r})=n_0\exp[\mp\psi({\bf r})]$, the PB theory
is based on the Poisson-Boltzmann equation
\begin{equation}
\nabla^2\psi=\kappa_0^2\sinh\psi~,
\label{PB}
\end{equation}
where $\psi=\beta e\phi$ is the reduced potential (vanishing in the reservoir),
$\beta=1/(k_BT)$ at temperature $T$, $\kappa_0=\sqrt{8\pi\lambda_B n_0}$ is the
Debye screening constant, and $\lambda_B=\beta e^2/\epsilon$ is the Bjerrum length.
Equation (\ref{PB}) must be solved together with appropriate boundary conditions:
$\nabla\psi|_{r=a}=Z_0\lambda_B/a^2$ and $\nabla\psi=0$ either as $r\to\infty$
--- far from the macroions in a bulk suspension --- or at $r=R$ in a symmetric cell
of radius $R$.
Neglecting macroion-macroion correlations, and all but asymptotically long-range
microion-microion correlations, the corresponding Helmholtz free energy takes the form
\begin{equation}
\beta F=\sum_{i=\pm}\int{\rm d}{\bf r}\,n_i({\bf r})[\ln(n_i({\bf r})\Lambda^3)-1]
+\frac{1}{8\pi\lambda_B}\int{\rm d}{\bf r}\,|\nabla\psi|^2~,
\label{FPB}
\end{equation}
where $\Lambda$ is the microion thermal de Broglie wavelength and the two terms
on the right side represent, respectively, the ideal-gas free energy due to
microion entropy and the total electrostatic energy.

At distances $r$ for which $|\psi(r)|\ll 1$, the right side of Eq.~(\ref{PB})
may be approximated by an expansion about the reservoir potential ($\psi=0$)
to linear order in $\psi$.
Anticipating applications to deionized suspensions of highly charged colloids,
however, the microion densities are here expanded instead about the mean (Donnan)
potential of the suspension $\bar\psi$~\cite{klein01,deserno02,tamashiro03}:
\begin{equation}
\nabla^2\psi=\kappa_0^2[\sinh\bar\psi+\cosh\bar\psi(\psi-\bar\psi)]~.
\label{LPB2}
\end{equation}
In a bulk suspension of macroions with bare valence $Z_0$, the solution of
Eq.~(\ref{LPB2}), with boundary condition $\psi'(r)\to 0$ as $r\to\infty$,
yields the potential generated by a single bare macroion~\cite{note1}:
\begin{equation}
\psi(r)=-Z_0\lambda_B~\frac{e^{\kappa a}}{1+\kappa a}
~\frac{e^{-\kappa r}}{r}~, \quad r\ge a~,
\label{psi0}
\end{equation}
with the bare screening constant
\begin{equation}
\kappa=\kappa_0\sqrt{\cosh\bar\psi}=\sqrt{4\pi\lambda_B(n_++n_-)}~.
\label{kappa}
\end{equation}
Here $n_{\pm}=n_0\exp(\mp\bar\psi)=N_{\pm}/[V(1-\eta)]$ represent the mean
number densities of microions in the {\it free} volume, i.e., the total volume $V$
reduced by the fraction $\eta$ occupied by the macroion hard cores.  Note that the
screening constant $\kappa$ depends implicitly on the average density of macroions,
since the global constraint of electroneutrality relates the numbers of macroions
($N_m$) and microions ($N_{\pm}$) in the suspension via $Z_0N_m=N_+-N_-$.
Combining the linearized PB equation [Eq.~(\ref{LPB2})] with a quadratic expansion
of the ideal-gas free energy [Eq.~(\ref{FPB})] about the mean microion densities
yields the corresponding linear-screening approximation for the one-body part
of the free energy per macroion:
\begin{equation}
\beta f=\sum_{i=\pm}x_i[\ln(n_i\Lambda^3)-1]
-\frac{Z_0^2}{2}\frac{\kappa\lambda_B}{1+\kappa a}
-\frac{Z_0^2}{2}~\frac{n_m}{n_++n_-}~,
\label{Ffree1}
\end{equation}
where $x_i=N_i/N_m$ and the three terms on the right side account for, respectively,
the microion entropy, macroion self-energy, and the Donnan potential energy of the
microions.

The PB theory proves to be formally equivalent to a class of effective-interaction
theories that map the macroion-microion mixture onto a one-component model (OCM),
by integrating over microion degrees of freedom in the partition function, and
that neglect all but long-range microion correlations~\cite{zvelindovsky07,denton07}.
The effective-interaction approach has been variously formulated as
density-functional~\cite{vRH97,graf98,vRDH99,vRE99,zoetekouw_pre06}, extended
Debye-H\"uckel~\cite{warren00}, distribution-function~\cite{chan85,chan01},
and response~\cite{silbert91,denton99,denton00,denton04,denton06} theories --- all
fundamentally equivalent~\cite{zvelindovsky07}, aside from technical differences in
the incorporation of excluded-volume effects.  The OCM is governed by an effective
Hamiltonian comprising a one-body volume energy and summations over pair and,
in general, many-body effective interactions.

Linearizing the PB equation about the mean potential [Eq.~(\ref{LPB2})], and the PB
free energy about the mean microion densities, is completely equivalent in the OCM
to linearizing the microion free energy about a reference system of neutral macroions
embedded in an electroneutral microion plasma~\cite{denton07} and neglecting many-body
effective interactions.  Furthermore, the volume energy in the OCM turns out to be
identical to the linearized PB free energy [Eq.~(\ref{Ffree1})].  A significant advantage
of the OCM, however, is its natural incorporation of effective interactions between
macroions.  An additional contribution to the total free energy then comes from the
effective (reduced) macroion-macroion pair potential~\cite{denton99,denton00}
\begin{equation}
\beta v_{\rm eff}(r)=Z_0^2\lambda_B\left(\frac{e^{\kappa a}}
{1+\kappa a}\right)^2\frac{e^{-\kappa r}}{r}~,\quad r>2a~.
\label{veff}
\end{equation}

Further progress requires uniting the OCM-based linear-screening theory of
charged colloids with the charge renormalization model of Sec.~\ref{model}.
To this end, the total free energy is first separated, according to
\begin{equation}
F=F_{\rm free}+F_{\rm bound}+F_m~,
\label{Ftot}
\end{equation}
into contributions from free and bound microions and from effective interactions
between macroions, respectively.
The linear-screening theory is then applied only to the free microions,
whose free energy per macroion is approximated by [cf.~Eq.~(\ref{Ffree1})]
\begin{equation}
\beta f_{\rm free}=\sum_{i=\pm}\tilde x_i[\ln(\tilde n_i\Lambda^3)-1]
-\frac{Z^2}{2}~\frac{\tilde\kappa\lambda_B}{1+\tilde\kappa(a+\delta)}
-\frac{Z^2}{2}~\frac{n_m}{\tilde n_++\tilde n_-}~,
\label{Ffree2}
\end{equation}
where $\tilde x_{\pm}=\tilde N_{\pm}/N_m$, $\tilde N_{\pm}$ are the numbers
of free microions, $\tilde n_{\pm}=\tilde N_{\pm}/[V(1-\tilde\eta)]$, and
$\tilde\eta=\eta(1+\delta/a)^3$ is the effective volume fraction of the
dressed macroions.
Generalizing Eq.~(\ref{psi0}), the electrostatic potential around a dressed
macroion of effective valence $Z$ and effective radius $a+\delta$ is given by
\begin{equation}
\tilde\psi(r)=-Z\lambda_B~\frac{e^{\tilde\kappa(a+\delta)}}{1+\tilde\kappa(a+\delta)}
~\frac{e^{-\tilde\kappa r}}{r}, \quad r\ge a+\delta~,
\label{psir}
\end{equation}
with a renormalized screening constant
\begin{equation}
\tilde\kappa=\sqrt{4\pi\lambda_B(\tilde n_++\tilde n_-)}~.
\label{kappar}
\end{equation}
The association shell thickness is now specified by combining Eqs.~(\ref{delta1})
and (\ref{psir}), yielding
\begin{equation}
\frac{Z\lambda_B}{[1+\tilde\kappa(a+\delta)](a+\delta)}=C~,
\label{delta2}
\end{equation}
and solving self-consistently for $\delta$ (given $Z$), noting that
$\tilde\kappa$ depends implicitly on $\delta$.

The free energy of the bound counterions decomposes naturally into entropic
and energetic contributions.
The first contribution is the ideal-gas free energy of the bound counterions,
given exactly by
\begin{equation}
\beta F_{\rm id}=4\pi N_m\int_a^{a+\delta}{\rm d}r\, r^2 n_b(r)
\left[\ln\left(n_b(r)\Lambda^3\right)-1\right]~,
\label{Fid1}
\end{equation}
where $n_b(r)$ is the number density profile of bound counterions within
the association shell and the integral covers the volume of the shell from
inner radius $a$ to outer radius $a+\delta$.
Although $n_b(r)$ could be obtained by solving the nonlinear PB equation [Eq.~(\ref{PB})]
(as in ref.~\cite{zoetekouw_prl06}), the present study explores a simpler approximation,
$\ln\left(n_b(r)\Lambda^3\right)\simeq\ln\left(n_b\Lambda^3\right)$, which yields
\begin{equation}
\beta F_{\rm id}\simeq
N_m(Z_0-Z)\left[\ln\left(n_b\Lambda^3\right)-1\right]~,
\label{Fid2}
\end{equation}
where $n_b=(Z_0-Z)/v_s$ is the mean density of bound counterions in the
association shell of volume $v_s=(4\pi/3)[(a+\delta)^3-a^3]$.
The second contribution to the bound counterion free energy is the electrostatic
energy $F_{\rm el}$ required to assemble the total charge of the dressed macroions
--- bare and bound charge --- from infinity.  An exact calculation would again
require knowledge of the bound counterion density profile.  Here we simply assume
$n_b(r)$ to be sharply peaked near $r=a$ and take
\begin{equation}
\beta F_{\rm el}\simeq
N_m\frac{Z^2\lambda_B}{2a}~.
\label{Fel}
\end{equation}
In the case of macroion charges below the renormalization threshold, $Z=Z_0$ and
$F_{\rm el}$ is a trivial constant that is irrelevant for thermodynamics.
At charges high enough that $Z<Z_0$, however, $F_{\rm el}$ becomes significant,
since $Z$ is state-dependent (as seen below).
Combining Eqs.~(\ref{Fid2}) and (\ref{Fel}), the bound-counterion free energy
per macroion is here simply approximated by
\begin{equation}
\beta f_{\rm bound}\simeq (Z_0-Z)\left[\ln\left(\frac{Z_0-Z}{v_s}\Lambda^3\right)-1\right]
+\frac{Z^2\lambda_B}{2a}~.
\label{Fb2}
\end{equation}

For a given bare valence $Z_0$, the effective valence $Z$ is prescribed
by minimizing with respect to $Z$ the total microion free energy
[sum of Eqs.~(\ref{Ffree2}) and (\ref{Fb2})] at fixed temperature and
mean microion densities:
\begin{equation}
\left(\frac{\partial}{\partial Z}(f_{\rm free}+f_{\rm bound})
\right)_{T,n_{\pm}}=0~.
\label{Z}
\end{equation}
The same variational prescription has been adopted by
Levin {\it et al.}~\cite{levin98,tamashiro98,levin01}.
It is easily shown that the minimization condition is equivalent to equating the
chemical potentials of counterions in the free and bound phases,
under the constraint that $Z$ and $\delta$ are related by Eq.~(\ref{delta2}).
The effective valence and corresponding shell thickness in turn determine the
effective screening constant $\tilde\kappa$ via Eq.~(\ref{kappar}).

Once the effective valence and screening constant are determined, the effective
pair potential between dressed macroions follows as
\begin{equation}
\beta\tilde v_{\rm eff}(r)=Z^2\lambda_B\left(\frac{e^{\tilde\kappa a}}
{1+\tilde\kappa a}\right)^2\frac{e^{-\tilde\kappa r}}{r}~,\quad r>2(a+\delta)~,
\label{veffr}
\end{equation}
from which the macroion free energy $F_m$ can be computed via liquid-state theory
or computer simulation.  (Note that the macroion radius is not renormalized in
the prefactor of the effective pair potential in Eq.~(\ref{veffr}), since the
association shells are penetrable.)
For demonstration purposes, we implement a variational method~\cite{vRH97,denton06}
based on first-order thermodynamic perturbation theory with a hard-sphere
reference system~\cite{HM}.
The macroion free energy per macroion is thus approximated as
\begin{equation}
\hspace*{-2.5cm}
f_m(n_m,\tilde n_{\pm})=\min_{(d)}\left\{f_{\rm HS}(n_m,\tilde n_{\pm};d)
+2\pi n_m\int_d^{\infty}{\rm d}r\,r^2g_{\rm HS}(r,n_m;d)
\tilde v_{\rm eff}(r,n_m,\tilde n_{\pm})\right\}~,
\label{fm}
\end{equation}
where the effective hard-sphere diameter $d$ is the variational parameter
and $f_{\rm HS}$ and $g_{\rm HS}$ are the excess free energy density and
(radial) pair distribution function, respectively, of the HS fluid,
computed here from the near-exact Carnahan-Starling and Verlet-Weis
expressions~\cite{HM}.  Minimization of $f_m$ with respect to $d$
generates a least upper bound to the free energy.
It is important, in practice, to fix the renormalized system parameters
($Z$, $\delta$, $\tilde\kappa$) in this minimization and in all partial
thermodynamic derivatives.

The thermodynamic pressure finally can be calculated from
\begin{equation}
p=n_m^2\left(\frac{\partial f}{\partial n_m}\right)_{T,N_s/N_m}
=p_{\rm free}+p_m~,
\label{ptot}
\end{equation}
where $f=F/N_m$ is the total Helmholtz free energy per macroion,
$N_s=N_-$ is the number of salt ion pairs in the suspension,
\begin{equation}
\beta p_{\rm free}=\tilde n_++\tilde n_-
-\frac{Z(\tilde n_+-\tilde n_-)\tilde\kappa\lambda_B}{4[1+\tilde\kappa(a+\delta)]^2}
\label{pfree}
\end{equation}
is the (reduced) pressure generated by the free microions, and
\begin{equation}
\beta p_m=n_m+n_m^2\beta\left(\frac{\partial f_m}{\partial n_m}\right)_{T,N_s/N_m}
\label{pm}
\end{equation}
is the macroion pressure due to macroion entropy and effective pair interactions.
Note that, since $Z$ and $\delta$ are implicitly held fixed in the partial
derivatives, the bound counterions make no contribution to the pressure.
As an alternative to variational theory, computer simulation also can be used
to determine the macroion pressure~\cite{lu-denton07}.

\section{Results and Discussion}\label{results}

To demonstrate its implementation, the charge renormalization theory is now applied
to deionized suspensions of charged colloids and monovalent microions in an aqueous
solvent at room temperature ($\lambda_B=0.72$ nm).
As noted in Sec.~\ref{theory}, the theory involves a single free parameter,
namely the dimensionless parameter $C$ in Eq.~(\ref{Z}), which establishes
the threshold for charge renormalization.  To ensure that a counterion's average
thermal energy does not exceed its binding potential, $C$ must be ${\cal O}(1)$.
Lacking an independent, physical criterion, $C$ must be regarded for the present
as a fitting parameter.  All results presented below were computed for $C=3$,
a value found to give satisfactory overall agreement with thermodynamic and structural
data from primitive model simulations.  In passing, we note that the thermal parameter
$C$ in the present theory is somewhat analogous to the adjustable cell radius
parameter $b$ in ref.~\cite{zoetekouw_prl06}, which combines PB cell and
linear-screening theories.

\begin{figure}[h!]
\begin{center}
\includegraphics[width=0.5\columnwidth]{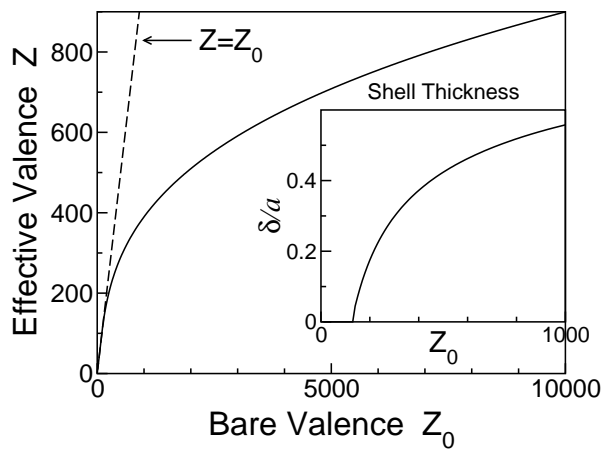}
\end{center}
\vspace*{-0.5cm}
\caption{\label{fig-crz}
Effective valence $Z$ vs.~bare valence $Z_0$ for a deionized suspension
($c_s\simeq 0$) of macroions of radius $a=50$ nm and volume fraction $\eta=0.1$.
Inset: counterion association shell emerges and thickens beyond threshold $Z_0$.
}
\end{figure}

\begin{figure}
\begin{center}
\includegraphics[width=0.5\columnwidth]{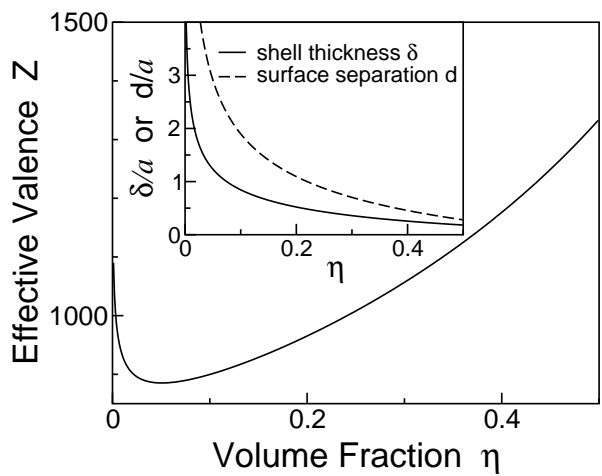}
\end{center}
\vspace*{-0.5cm}
\caption{\label{fig-cre}
Effective valence $Z$ vs.~volume fraction $\eta$ for a deionized suspension
of macroions of radius $a=50$ nm and bare valence $Z_0=10^4$.
Inset: association shell (solid curve) thins with increasing $\eta$,
remaining thinner than nearest-neighbour surface separation in fcc crystal
(dashed curve).
}
\end{figure}

\begin{figure}
\begin{center}
\includegraphics[width=0.5\columnwidth]{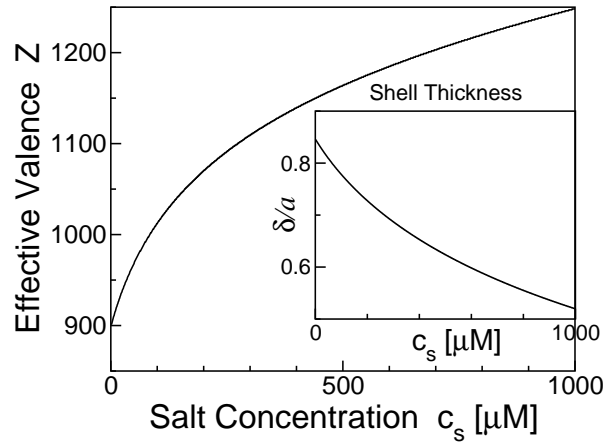}
\end{center}
\vspace*{-0.5cm}
\caption{\label{fig-crzds}
Effective valence $Z$ vs.~system salt concentration $c_s$ for macroions of
radius $a=50$ nm, bare valence $Z_0=10^4$, and volume fraction $\eta=0.1$.
Inset: association shell thins with increasing salt concentration.
}
\end{figure}

The key physical concepts of the charge renormalization theory are illustrated
in Figs.~\ref{fig-crz}-\ref{fig-crzds}.  For a sufficiently small bare valence,
Eq.~(\ref{delta2}) admits no real solution for nonzero thickness of the association
shell.  In this case, there are no bound counterions ($\delta=0$, $v_s=0$)
and the free energy is minimized by $Z=Z_0$ (dashed line in Fig.~\ref{fig-crz}).
At a threshold value of the bare valence, however, the shell emerges continuously
and thickens rapidly with increasing $Z_0$ at fixed volume fraction and salt
concentration (inset to Fig.~\ref{fig-crz}), while the free energy minimum shifts
to $Z<Z_0$ (solid curve in Fig.~\ref{fig-crz}).
The effective valence does not saturate, but continues to grow logarithmically
with increasing $Z_0$, in contrast to the behaviour expected and observed for
polyelectrolytes~\cite{Oosawa} and to predictions for colloidal suspensions
from PB cell-based theories~\cite{alexander84,belloni98,trizac02} and
Debye-H\"uckel-based theories~\cite{levin98,tamashiro98,levin01,levin03,trizac04}.

As Figs.~\ref{fig-cre} and \ref{fig-crzds} show, $Z$ varies with volume fraction
$\eta$ and salt concentration $c_s$ of the suspension.  The effective valence
thus depends nontrivially on the thermodynamic state, exhibiting a pronounced
minimum with respect to $\eta$ and increasing monotonically with $c_s$.
Correspondingly, the association shell thins with increasing $\eta$ and $c_s$
(Figs.~\ref{fig-cre} and \ref{fig-crzds}, insets), in such a manner, however,
that the shells surrounding neighbouring macroions always remain separate and distinct.
The absence of overlapping shells provides an internal consistency check on the theory.

\begin{figure}
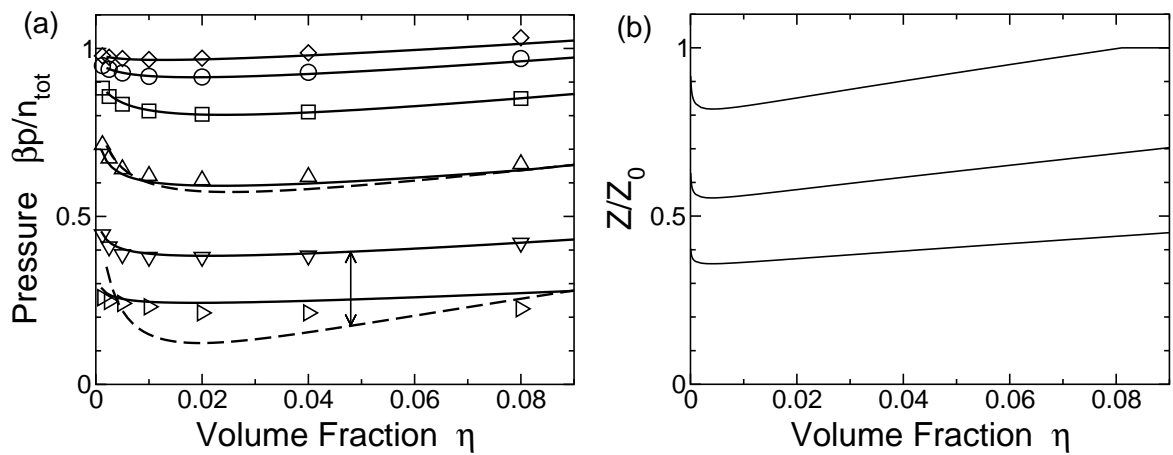

\begin{center}
\includegraphics[width=0.48\columnwidth]{p-vs-linse.z40.cr.eps}
\hspace*{0.1cm}
\includegraphics[width=0.48\columnwidth]{zcr.z40.eps}
\end{center}
\vspace*{-0.5cm}
\caption{\label{fig-crpe}
(a) Total reduced pressure $\beta p/n_{\rm tot}$ vs.~macroion volume fraction $\eta$,
where $n_{\rm tot}=(Z_0+1)n_m$ (total ion density), of salt-free suspensions
with bare macroion valence $Z_0=40$ and electrostatic coupling constants
(top to bottom) $\Gamma=0.0222$, 0.0445, 0.0889, 0.1779, 0.3558, 0.7115.
Symbols: Monte Carlo simulations of the primitive model~\cite{linse00}
(symbol sizes exceed error bars).  Curves: variational theory with
(solid) and without (dashed) charge renormalization.
The double-ended arrow points to corresponding curves for $\Gamma=0.3558$.
The dashed curve for $\Gamma=0.7115$ is off-scale, the pressure being negative.
(b) Corresponding ratio of effective to bare macroion valence $Z/Z_0$ vs. $\eta$
for $\Gamma=0.1779$, 0.3558, 0.7115 (top to bottom).  For $\Gamma\le 0.1$,
no renormalization is predicted ($Z=Z_0$).
}
\end{figure}

\begin{figure}
\begin{center}
\vspace*{1cm}
\includegraphics[width=0.5\columnwidth]{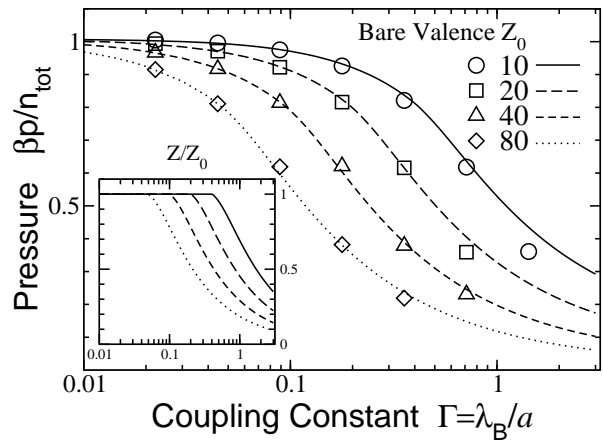}
\end{center}
\vspace*{-0.5cm}
\caption{\label{fig-crpg}
Total reduced pressure $\beta p/n_{\rm tot}$ vs.~electrostatic coupling
constant $\Gamma$ of salt-free suspensions with fixed volume fraction $\eta=0.01$
and bare macroion valence (top to bottom) $Z_0=10, 20, 40, 80$.
Open symbols: Monte Carlo simulations of the primitive model~\cite{linse00}
(symbol sizes exceed error bars).  Curves: charge-renormalized variational theory.
Inset: Ratio of effective to bare macroion valence $Z/Z_0$ vs. $\Gamma$.
}
\end{figure}

To test predictions of the theory for thermodynamic properties, the pressures of
deionized suspensions, calculated from Eqs.~(\ref{kappar}), (\ref{delta2}), and
(\ref{ptot})-(\ref{pm}), are directly compared with available data from simulations
of the primitive model.  Figures~\ref{fig-crpe} and \ref{fig-crpg} show comparisons
with the results of Linse~\cite{linse00} from extensive Monte Carlo simulations of
salt-free suspensions with various bare valences and electrostatic coupling parameters
$\Gamma=\lambda_B/a$.  The unrenormalized linear-screening theory~\cite{denton99,denton00}
performs excellently for low-to-moderate couplings ($\Gamma<0.1779$ for $Z_0=40$
in Fig.~\ref{fig-crpe}), but breaks down at higher couplings, characteristic of
highly charged latex particles and ionic surfactant micelles.
As illustrated in Figs.~\ref{fig-crpe}(b) and \ref{fig-crpg} (inset), charge
renormalization becomes important for $Z_0\Gamma>7$, where the effective valence
tends to be lower than the bare valence.  The renormalized theory restores close
agreement with simulation up to at least $Z_0\Gamma\simeq 28$ ($\Gamma=0.7115$
in Fig.~\ref{fig-crpe}).  In practice, the excluded-volume correction to the microion
densities in Eq.~(\ref{Ffree2}), and the inclusion of the effective pair pressure
--- already important features of the unrenormalized theory~\cite{lu-denton07} ---
are essential for consistent quantitative accuracy as the volume fraction becomes
renormalized.  Remarkably and intriguingly, the threshold for charge renormalization
coincides with the onset of a spinodal phase instability, at low but non-zero salt
concentrations, predicted by linear-screening
theories~\cite{vRH97,vRDH99,warren00,denton06}.  This rather unusual prediction,
however, has not yet been confirmed by primitive model simulations and the
experimental situation is unresolved.

\begin{figure}
\begin{center}
\vspace*{1cm}
\includegraphics[width=0.5\columnwidth]{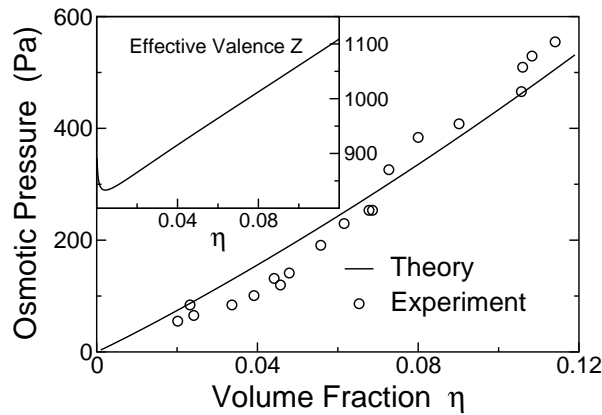}
\end{center}
\vspace*{-0.5cm}
\caption{\label{fig-reus}
Osmotic pressure (in Pa units) vs.~volume fraction $\eta$ for a deionized
suspension ($c_s=0$) of charged macroions of radius $a=51$ nm and
bare valence $Z_0=3000$.  Curve: charge renormalization theory; symbols:
experimental data~\cite{reus97}.  Inset: effective valence $Z$ vs.~$\eta$.
}
\end{figure}

Finally, the theory also can be tested against available experimental data.  Figure
\ref{fig-reus} shows a comparison of predictions with osmotic pressure measurements
of deionized, aqueous, charged colloidal crystals reported by
Reus {\it et al.}~\cite{reus97}.  Since bare (titratable) charges are notoriously
difficult to characterize in experiments,
$Z_0$ is treated here as a fitting parameter, a value of $Z_0\simeq 3000$ giving a
reasonable fit to the data.  As seen in the inset to Fig.~\ref{fig-reus}, the effective
charge in this case is substantially lower than the bare charge.  It is important to
emphasize, however, that only thermodynamic quantities have physical significance
within the theory and that the theoretically defined variable $Z$ does not necessarily
correspond directly with any effective charge that may be determined experimentally,
e.g., by light scattering, electrophoresis, or conductivity measurements.
Moreover, direct comparisons between theory and experiment are subject to complication
by charge regulation via chemical reactions at the macroion
surface~\cite{belloni91,vongrunberg99}, which may render even the bare charge
dependent on thermodynamic state, e.g., pH and salinity.
\newpage

\section{Conclusions}\label{conclusions}

In summary, a new theory of charge renormalization in charge-stabilised colloidal
suspensions has been developed and implemented.  The theory posits the existence of
free and bound counterion phases and integrates a thermal criterion for distinguishing
between the two phases with an effective-interaction theory based on a one-component
model.  Within the theory, bound counterions act to renormalize the effective valence
of the dressed macroions, while free counterions screen the dressed macroions and
make the dominant contribution to the pressure.  A linear-screening approximation
accurately describes monovalent free counterions, while the bound counterions are
adequately described by a comparatively crude coarse-grained approximation for
the bound-counterion density profile.

Despite the conceptual and practical simplicity of the charge renormalization theory,
predictions for the pressure closely agree with corresponding data from both
primitive model simulations and an experiment, over ranges of macroion charges,
volume fractions, and electrostatic coupling strengths, demonstrating the practical
potential of the theory for modeling equilibrium thermodynamic properties.
A preliminary simulation study~\cite{lu-denton08} indicates that the theory also can
accurately model structural properties, such as macroion-macroion pair distribution
functions, although within a more limited range of electrostatic couplings.
Future work will focus on refinements of the theory, further comparisons with
experiment, and applications to the phase behaviour of deionized suspensions of
highly charged macroions in bulk and
in confinement~\cite{dietrich07,zwannikken07,klapp08}.

\ack
This work was supported by the National Science Foundation (DMR-0204020)
and the Petroleum Research Fund (PRF 44365-AC7).  Fruitful discussions with
Ben Lu and helpful correspondence with Per Linse are gratefully acknowledged.



\end{document}